\documentstyle[12pt,epsf,epsfig]{article}
\newcount\timecount
\newcount\hours \newcount\minutes  \newcount\temp \newcount\pmhours

\hours = \time
\divide\hours by 60
\temp = \hours
\multiply\temp by 60
\minutes = \time
\advance\minutes by -\temp
\def\hour{\the\hours}
\def\minute{\ifnum\minutes<10 0\the\minutes
            \else\the\minutes\fi}
\def\clock{
\ifnum\hours=0 12:\minute\ AM
\else\ifnum\hours<12 \hour:\minute\ AM
      \else\ifnum\hours=12 12:\minute\ PM
            \else\ifnum\hours>12
                 \pmhours=\hours
                 \advance\pmhours by -12
                 \the\pmhours:\minute\ PM
                 \fi
            \fi
      \fi
\fi
}

\def\monthname{\relax\ifcase\month 0/\or January\or February\or
   March\or April\or May\or June\or July\or August\or September\or
   October\or November\or December\else\number\month/\fi}

\def\bold#1{\setbox0=\hbox{$#1$}%
     \kern-.025em\copy0\kern-\wd0
     \kern.05em\copy0\kern-\wd0
     \kern-.025em\raise.0433em\box0 }

\def\ga{\mathrel{\raise.3ex\hbox{$>$\kern-.75em\lower1ex\hbox{$\sim$}}}}
\def\la{\mathrel{\raise.3ex\hbox{$<$\kern-.75em\lower1ex\hbox{$\sim$}}}}
\def\gev{{\rm \, Ge\kern-0.125em V}}
\def\tev{{\rm \, Te\kern-0.125em V}}
\def\beq{\begin{equation}}
\def\eeq{\end{equation}}

\def\m12{m_{1\!/2}}

\def\ga{\mathrel{\raise.3ex\hbox{$>$\kern-.75em\lower1ex\hbox{$\sim$}}}}
\def\la{\mathrel{\raise.3ex\hbox{$<$\kern-.75em\lower1ex\hbox{$\sim$}}}}
\def\gyr{{\rm \, G\kern-0.125em yr}}
\def\gev{{\rm \, Ge\kern-0.125em V}}
\def\tev{{\rm \, Te\kern-0.125em V}}
\def\beq{\begin{equation}}
\def\eeq{\end{equation}}

\def\stau{\tilde \tau}

\def\m12{m_{1\!/2}}

\def\gappeq{\mathrel{\rlap {\raise.5ex\hbox{$>$}}
{\lower.5ex\hbox{$\sim$}}}}

\def\lappeq{\mathrel{\rlap{\raise.5ex\hbox{$<$}}
{\lower.5ex\hbox{$\sim$}}}}

\def\Toprel#1\over#2{\mathrel{\mathop{#2}\limits^{#1}}}

\begin{document}
\begin{titlepage}
\pagestyle{empty}
\baselineskip=21pt
\rightline{hep-ph/0204192}
\rightline{CERN--TH/2002-081}
\rightline{UMN--TH--2049/02}
\rightline{TPI--MINN--02/09}
\vskip 1in
\begin{center}
{\large{\bf The MSSM Parameter Space with Non-Universal Higgs Masses}}
\end{center}
\begin{center}
\vskip 0.2in
{{\bf John Ellis}$^1$,  {\bf Keith
A.~Olive}$^{2}$ and {\bf Yudi Santoso}$^{2}$}\\
\vskip 0.1in
{\it
$^1${TH Division, CERN, Geneva, Switzerland}\\
$^2${Theoretical Physics Institute,
University of Minnesota, Minneapolis, MN 55455, USA}}\\
\vskip 0.2in
{\bf Abstract}
\end{center}
\baselineskip=18pt \noindent


Without assuming that Higgs masses have the same values as other scalar
masses at the input GUT scale, we combine constraints on the minimal
supersymmetric extension of the Standard Model (MSSM) coming from the cold
dark matter density with the limits from direct searches at accelerators
such as LEP, indirect measurements such as $b \to s \gamma$ decay and the
anomalous magnetic moment of the muon. The requirement that Higgs
masses-squared be positive at the GUT scale imposes important restrictions
on the MSSM parameter space, as does the requirement that the LSP be
neutral. We analyze the interplay of these constraints in the $(\mu,
m_A)$, $(\mu, m_{1/2}), (m_{1/2}, m_0)$ and $(m_A, \tan \beta)$ planes.
These exhibit new features not seen in the corresponding planes in the
constrained MSSM in which universality is extended to Higgs masses.

\vfill
\leftline{CERN--TH/2002-081}
\end{titlepage}

\section{Introduction}

In order to avoid fine tuning to preserve the mass hierarchy $m_W \ll
m_P$~\cite{hierarchy}, supersymmetry at the TeV scale is commonly
postulated. Cosmology also favours the TeV mass range, if the lightest
supersymmetric particle (LSP) is stable, as occurs if $R$ parity is
conserved. In the following, we work within the minimal supersymmetric
extension of the Standard Model (MSSM). Furthermore, in order to satisfy
the strong constraints on charged or colored dark matter\cite{isotopes},
we require that the LSP is neutral. Over almost all of the MSSM parameter
space the lightest neutral sparticle is a neutralino\footnote{We
comment below on specific cases where the sneutrino may be the LSP.}
$\chi$, i.e., a mixture of the
$\tilde B, \tilde W^3, \tilde H_1$ and
$\tilde H_2$. In this case, the LSP would be an excellent candidate for
astrophysical dark matter~\cite{EHNOS}.

A key uncertainty in the MSSM is the pattern of soft supersymmetry
breaking, as described by the scalar masses $m_0$, gaugino masses
$m_{1/2}$ and trilinear couplings $A_0$~\cite{DG}. These presumably
originate from physics at some high-energy scale, e.g., from some
supergravity or superstring theory, and then evolve down to lower energy
scale according to well-known renormalization-group equations. What is
uncertain, however, is the extent to which universality applies to the
scalar masses
$m_0$ for different squark, slepton and Higgs fields, the gaugino masses
$m_{1/2}$ for the $SU(3)$, $SU(2)$ and $U(1)$ gauginos, and the trilinear
couplings $A$ corresponding to different Yukawa couplings. We do not
consider here non-universal gaugino masses or $A$ parameters. 

The suppression of flavour-changing neutral interactions~\cite{EN}
suggests that the $m_0$ may be universal for different matter fields with
the same quantum numbers, e.g., the different squark and slepton
generations~\cite{BG}.  However, there is no very good reason to postulate
universality between, say, the spartners of left- and right-handed quarks,
or between squarks and sleptons. In Grand Unified Theories (GUTs), there
must be universality between fields in the same GUT multiplet, e.g., $u_L,
d_L, u_R$ and $e_R$ in a ${\mathbf 10}$ of $SU(5)$, and this would extend
to all matter fields in a ${\mathbf 16}$ of $SO(10)$. However, there is
less reason to postulate universality between these and the Higgs
fields. Nevertheless, this extension of universality to the Higgs masses
(UHM) is often assumed, resulting in what is commonly termed the
constrained MSSM (CMSSM). Alternatively, there may be non-universal higgs
masses (NUHM) in the more general MSSM.

We and others have previously made extensive studies of the allowed
parameter space in the CMSSM \cite{EFGO,EFGOSi,otherOmega,coann,stopco},
incorporating experimental constraints and the requirement that the relic
density
$\Omega_\chi h^2$ fall within a range $0.1 < \Omega_\chi h^2 < 0.3$
preferred by cosmology. There have also been many studies of the NUHM case
in the MSSM~\cite{nonu,higgsino,EFGOS,EFGO}, in which the character of
the LSP may change, perhaps becoming mainly a Higgsino ${\tilde H}$, 
rather than
a Bino ${\tilde B}$ as in the CMSSM. We think it is opportune to study
the NUHM case again, taking into account more recent improvements in the
understanding of the cosmological relic density $\Omega_\chi h^2$,
including $\chi - {\tilde \tau}$~\cite{coann} and $\chi - {\tilde
t}$~\cite{stopco} coannihilations in addition to $\chi - \chi^\prime -
\chi^\pm$ coannihilations \cite{oldcoann} and the r\^oles of
direct-channel MSSM Higgs resonances~\cite{funnel,EFGOSi}, as well as the
(almost) final versions of the direct constraints imposed by LEP
experiments.

\section{Experimental Constraints and the NUHM Analysis}

Important experimental constraints on the MSSM parameter space are
provided by direct searches at LEP, such as that on the lightest chargino 
$\chi^\pm$: 
$m_{\chi^\pm} \gappeq$ 103.5 GeV~\cite{LEPsusy}, and that on the 
selectron $\tilde e$: $m_{\tilde e}\gappeq$ 99 GeV \cite{LEPSUSYWG_0101}, 
depending slightly on the other MSSM parameters. For our 
purposes, another
important constraint is provided by the LEP lower limit on the
Higgs mass: $m_H > 114$~GeV \cite{LEPHiggs} in the
Standard Model. This limit may be applicable to the lightest Higgs boson 
$h$ in the general MSSM, although possibly in relaxed form~\footnote{As we 
discuss later, there is no such relaxation in the regions of MSSM 
parameter space of interest to us.}. 
We recall that $m_h$ is sensitive to sparticle
masses, particularly $m_{\tilde t}$, via loop 
corrections~\cite{radcorrH,FeynHiggs}:
\beq
\delta m^2_h \propto {m^4_t\over m^2_W}~\ln\left({m^2_{\tilde t}\over
m^2_t}\right)~ + \ldots
\label{nine}
\eeq
which implies that the LEP Higgs limit constrains
the MSSM parameters.

We also impose the constraint imposed by measurements of $b\rightarrow
s\gamma$~\cite{bsg}, ${\rm BR}(B \rightarrow
X_s \gamma) = (3.11 \pm 0.42 \pm 0.21) \times 10^{-4}$, which agree with the
Standard Model calculation ${\rm BR}(B \rightarrow X_s
\gamma)_{\rm SM} = (3.29 \pm 0.33) \times 10^{-4}$~\cite{bsgSM}. Typically, the
$b\rightarrow s\gamma$ constraint is more important for $\mu < 0$, but it
is also relevant for $\mu > 0$, particularly when $\tan\beta$ is large.

We also take into account the anomalous magnetic moment of the muon. The
BNL E821~\cite{BNL} experiment reported a new measurement of $a_\mu\equiv
{1\over 2} (g_\mu -2)$ which deviates by 1.6 standard deviations from the
best Standard Model prediction (once the pseudoscalar-meson pole part of
the light-by-light scattering contribution~\cite{lightbylight} is
corrected).  Currently, the deviation from the Standard Model value is $-6
\times 10^{-10}$ to $58 \times 10^{-10}$ at the 2-$\sigma$ level. The
2-$\sigma$ limit still prefers~\cite{susygmu} the $\mu > 0$ part of
parameter space, but $\mu < 0 $ is allowed so long as either
(or both) $m_{1/2}$ and $m_0$ are large~\cite{EOS2}. Where appropriate,
the current
$g_\mu -2$ constraint is taken into account~\footnote{We note that a new
experimental value with significantly reduced statistical error is
expected soon, but the theoretical interpretation will still be subject to
strong-interaction uncertainties in the Standard Model prediction, so the
impact may be muffled.}.

In the following, we display the regions of MSSM parameter space where the 
supersymmetric
relic density $\rho_\chi \equiv \Omega_\chi \rho_{critical}$ falls within 
the
following preferred range:
\beq
0.1 < \Omega_\chi h^2 < 0.3.
\label{ten}
\eeq
The upper limit is rigorous, and assumes only that the age of the Universe
exceeds 12 Gyr. It is also consistent with
the total matter density $\Omega_m \lappeq 0.4$, and the Hubble expansion
rate $h \sim 1/\sqrt{2}$ to within about 10 \% (in units of 100 km/s/Mpc). 
An upper limit stronger than (\ref{ten}) could be defensible~\cite{MS}, in 
particular because global fits to cosmological data may favour a lower 
value of $\Omega_m$.  
On the other hand, the lower limit in (\ref{ten}) might be weakened, in
particular if there are other important contributions to the overall 
matter density.

In the CMSSM, the values of the Higgsino mixing parameter $\mu$ and 
the pseudoscalar Higgs mass $m_A$ are determined by the electroweak vacuum 
conditions, once $m_{1/2}, m_0, \tan \beta$ and the trilinear 
supersymmetry-breaking parameter $A_0$ are fixed. This is no longer the
case  when the universality assumption is relaxed for the Higgs
multiplets. No longer assuming that the Higgs soft masses $m_1$ and 
$m_2$~\footnote{The Higgs multiplets $H_{1,2}$ couple to
$d, u$ quarks, respectively.} are set
equal to $m_0$ at the GUT scale, we are free to choose $\mu$ and $m_A$ as
surrogate parameters. Thus we have as our free parameters $m_0$, $m_{1/2}$,
$A_0$,
$\tan \beta$, $\mu$ and $m_A$. The 
results depend somewhat on the masses of the top and bottom quarks that 
are assumed: for definiteness, we use the pole mass $m_t = 175$~GeV and 
the running mass $m_b (m_b)^{\overline{MS}}_{SM} = 4.25$~GeV.

As noted above, we include  $\chi - {\tilde l}$ ($l = e, \mu, \tau$)
\cite{coann} and
$\chi - {\tilde t}$ \cite{stopco} coannihilations in our calculation.
However, since we assume for simplicity  that $A_0 = 0$,  $\chi -
{\tilde t}$ coannihilation is not very important. Since $\mu$ is now a 
free parameter, the neutralino can be Bino-like or Higgsino-like depending
on the ratio of
$\mu$ to
$m_{1/2}$. When the LSP is Higgsino-like, it is often nearly
degenerate with the second lightest neutralino
$\chi^{\prime}$ and the lightest chargino $\chi^\pm$ \cite{oldcoann}. All
relevant coannihilation processes for this case, 
have been included. In some small regions of the parameter space $\chi$,
$\chi^{\prime}$,
$\chi^+$ and $\tilde{l}$ can all be degenerate, and we have also taken into
account the $\chi^{\prime} - {\tilde l}$ and $\chi^\pm - {\tilde l}$
coannihilations.

\section{Analysis of the $(\mu, m_A)$ Plane}

\begin{figure}
\vskip 0.5in
\vspace*{-0.75in}
\begin{minipage}{8in}
\epsfig{file=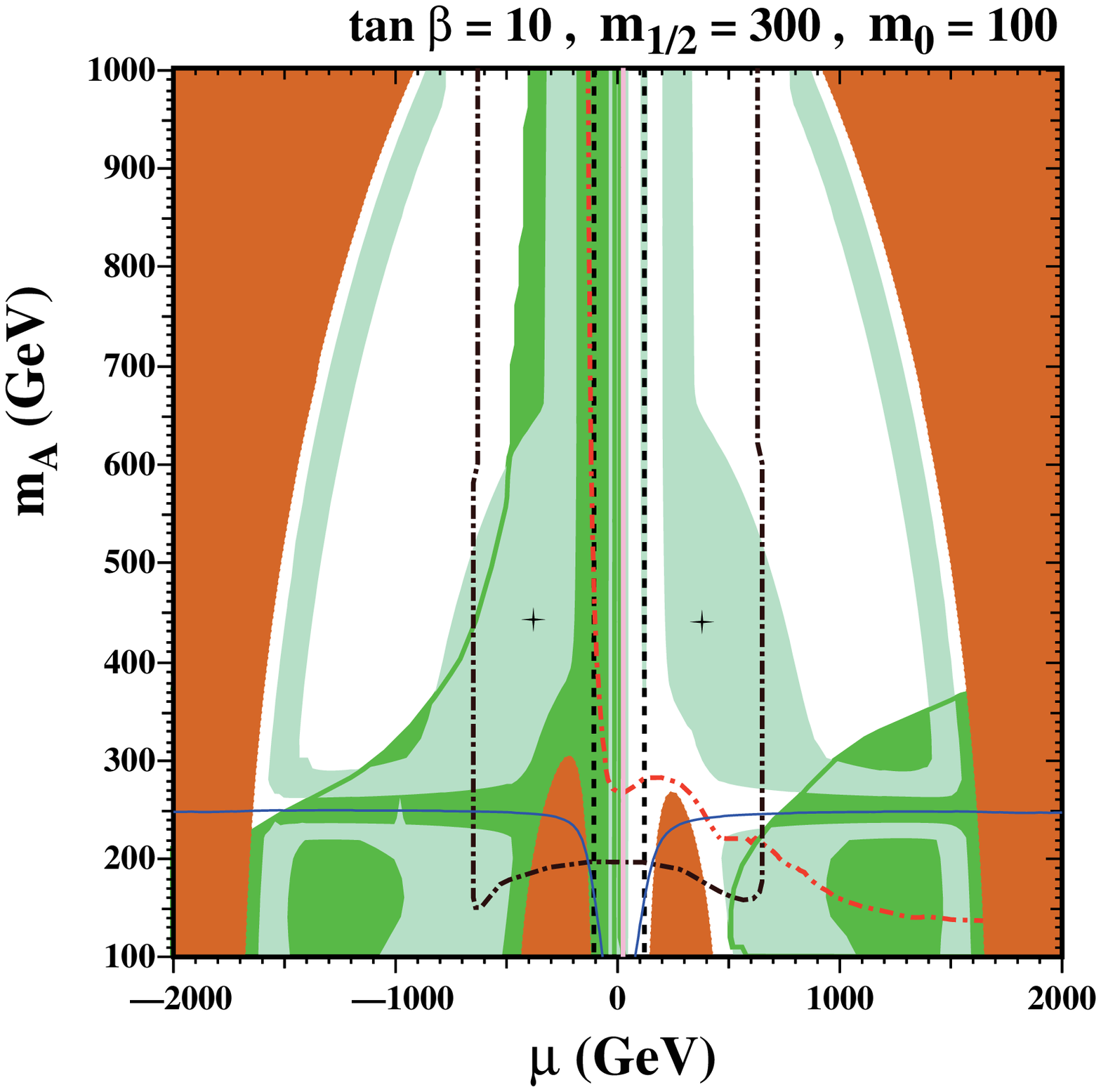,height=3.3in}
\hspace*{-0.17in}
\epsfig{file=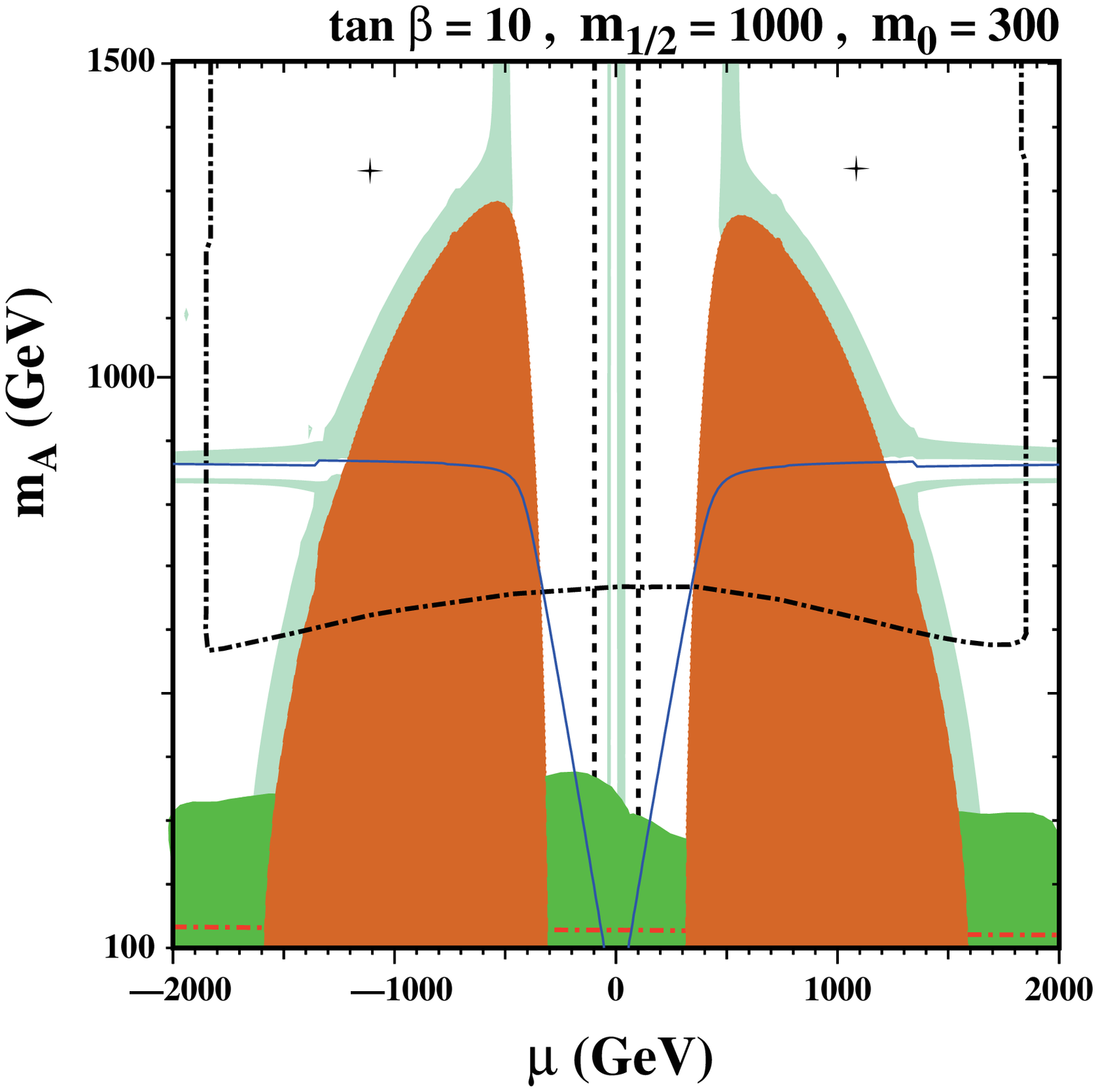,height=3.3in} \hfill
\end{minipage}
\caption{
{\it Compilations of phenomenological constraints on the MSSM with NUHM 
in the $(\mu, m_A)$ plane for $\tan \beta = 10$ and (a) $m_0 = 100$~GeV, 
$m_{1/2} = 300$~GeV, (b)  $m_0 = 300$~GeV,
$m_{1/2} = 1000$~GeV, assuming $A_0 = 0$, $m_t = 175$~GeV and 
$m_b(m_b)^{\overline {MS}}_{SM} = 4.25$~GeV.
The light (turquoise) shading denotes the region where $0.1 < \Omega_\chi 
h^2 < 0.3$, and 
the (blue) solid line is the contour $m_\chi = m_A/2$, near which
rapid direct-channel annihilation suppresses the relic density. 
The darker (green) shading shows the impact of the $b \to s \gamma$ 
constraint, and the darkest (red) shading shows where the LSP is charged.
The dark (black) dashed line is the chargino constraint $m_{\chi^\pm} > 
104$~GeV: lower $|\mu|$ values are not allowed.
The lighter (red) dot-dashed line is the contour $m_h = 114$~GeV 
calculated using {\tt FeynHiggs}~\cite{FeynHiggs}: lower $m_A$ values are 
not allowed.
The dark (black) dot-dashed line indicates when one or another Higgs 
mass-squared becomes negative at the GUT scale: only lower $|\mu|$ and 
larger $m_A$ values are allowed. The crosses denote the values of 
$\mu$ and $m_A$ found in the CMSSM.
}}
\label{fig:mumA10}
\end{figure}

We start our analysis by studying the range of  possibilities in the
$(\mu, m_A)$ plane for various fixed choices of 
$m_{1/2}$, $m_0$ and $\tan \beta$.
Panel (a) of Fig.~\ref{fig:mumA10} displays the $(\mu, m_A)$ plane for
$m_{1/2} = 300$~GeV, $m_0 = 100$~GeV and $\tan \beta = 10$. The very dark
(red) regions at large $|\mu|$ appear where the LSP is no longer the
neutral $\chi$ but the charged lighter stau ${\tilde \tau}_1$, which is
unacceptable astrophysically. There are also small `shark's teeth' at
$|\mu| \sim 400$~GeV, $m_A \lappeq 300$~GeV where the ${\tilde \tau}_1$ is
the LSP. At large $|\mu|$, the ${\tilde \tau}_1$ is driven light primarily
by the large mixing term in the stau mass matrix. At small $|\mu|$,
particularly at small $m_A$ when the mass difference $m_2^2 - m_1^2$ is
small, first the ${\tilde \tau}_R$ mass is driven small, making the
${\tilde \tau}_1$ the LSP again. However, at even smaller $|\mu|$ the
lightest neutralino gets lighter again, since $m_\chi \simeq \mu$ when
$\mu < M_1 \simeq 0.4 \, m_{1/2}$.

The light (turquoise) shaded region in panel (a) of Fig.~\ref{fig:mumA10}
is that for which $0.1 < \Omega_\chi h^2 < 0.3$. We see strips adjacent to
the ${\tilde \tau}_1$ LSP regions, where $\chi - {\tilde \tau}_1$
coannihilation \cite{coann} is important in suppressing the relic density
to an acceptable level. The thick cosmological region at smaller $\mu$
corresponds to the `bulk' region familiar from CMSSM studies. The two
(black) crosses indicate the position of the CMSSM points for these input
parameters.  Extending upward in $m_A$ from this region, there is another
light (turquoise) shaded band at smaller $|\mu|$. Here, the neutralino
gets more Higgsino-like and the annihilation to $W^+ W^-$ becomes
important, yielding a relic density in the allowed range~\footnote{This is
similar to the focus-point region \cite{focus} in the CMSSM.}. For smaller
$|\mu|$, the relic density becomes too small due to the $\chi -
\chi^{\prime} - \chi^+$ coannihilations. For even smaller $|\mu|$ ($
\stackrel{<}{\sim} 30$~GeV) many channels are kinematically unavailable
and we are no longer near the $h$ and $Z$ pole. As a result the relic
density may again come into the cosmologically preferred region. However,
this region is excluded by the LEP limit on the chargino mass as explained
below.

The unshaded regions between the allowed bands have a relic density that
is too high: $\Omega_\chi h^2 > 0.3$. However, the $\tilde \tau$
coannihilation and bulk bands are connected by horizontal bands of
acceptable relic density that are themselves separated by unshaded regions
of low relic density, threaded by solid (blue) lines asymptoting to $m_A
\sim 250$~GeV. These lines correspond to cases when $m_\chi \simeq m_A /
2$, where direct-channel annihilation: $\chi + \chi \to A,H$ is important,
and suppresses the relic density \cite{EFGOSi,funnel} creating
`funnel'-like regions.

Overlaying the cosmological regions are dark
(green)  shaded regions excluded by $b \to s \gamma$. That for $\mu < 0$
is more important, as expected from previous analyses. Taking this into
account, most of the bulk and coannihilation regions are allowed for $\mu
> 0$, but only the coannihilation regions and a small slice of the bulk
region for $\mu < 0$. In this example, the `funnel' regions are largely
excluded by $b \to s \gamma$ for both signs of $\mu$.

\begin{figure}
\vskip 0.5in
\vspace*{-0.75in}
\begin{minipage}{8in}
\epsfig{file=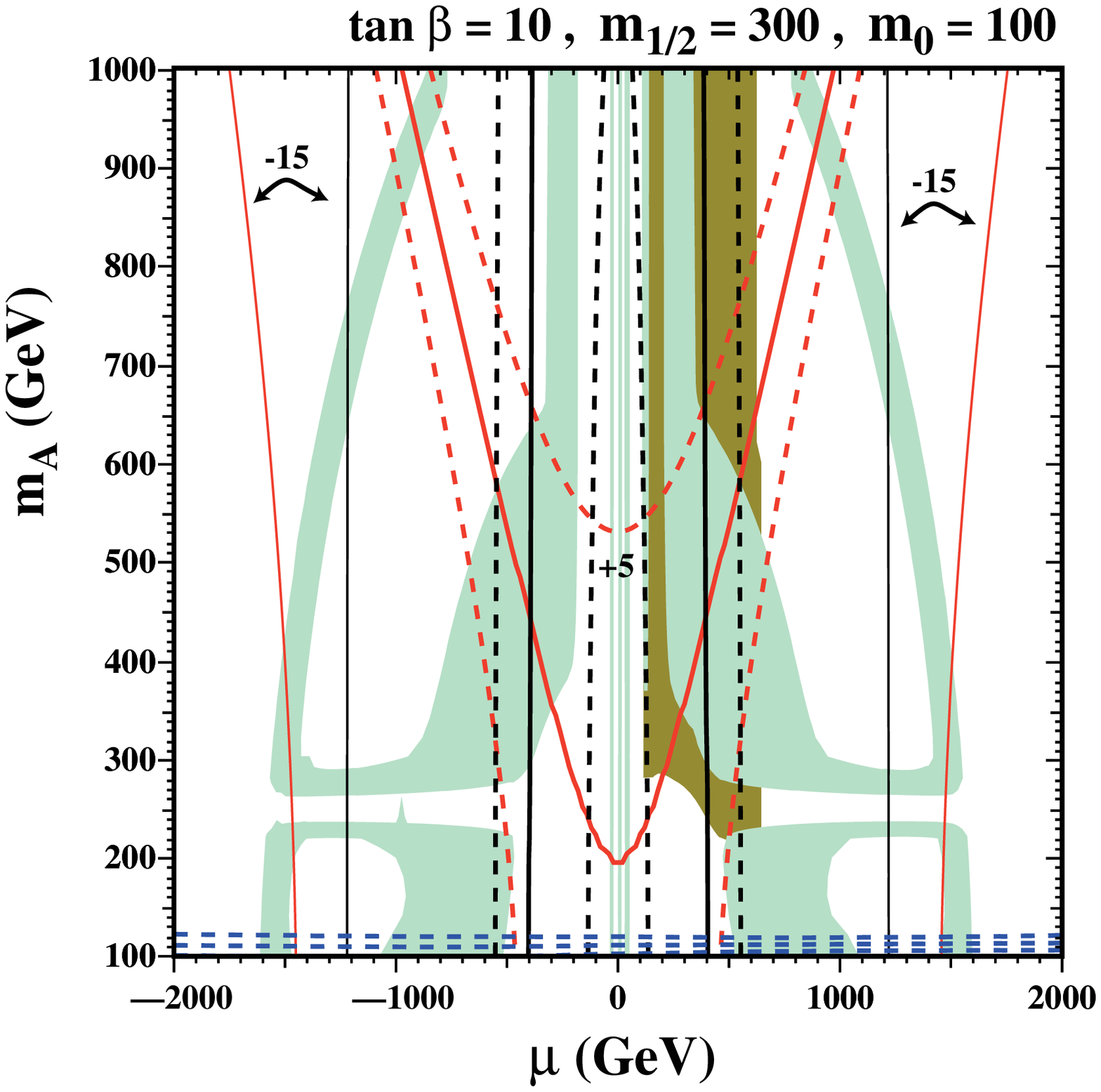,height=3.3in}
\hspace*{-0.17in}
\epsfig{file=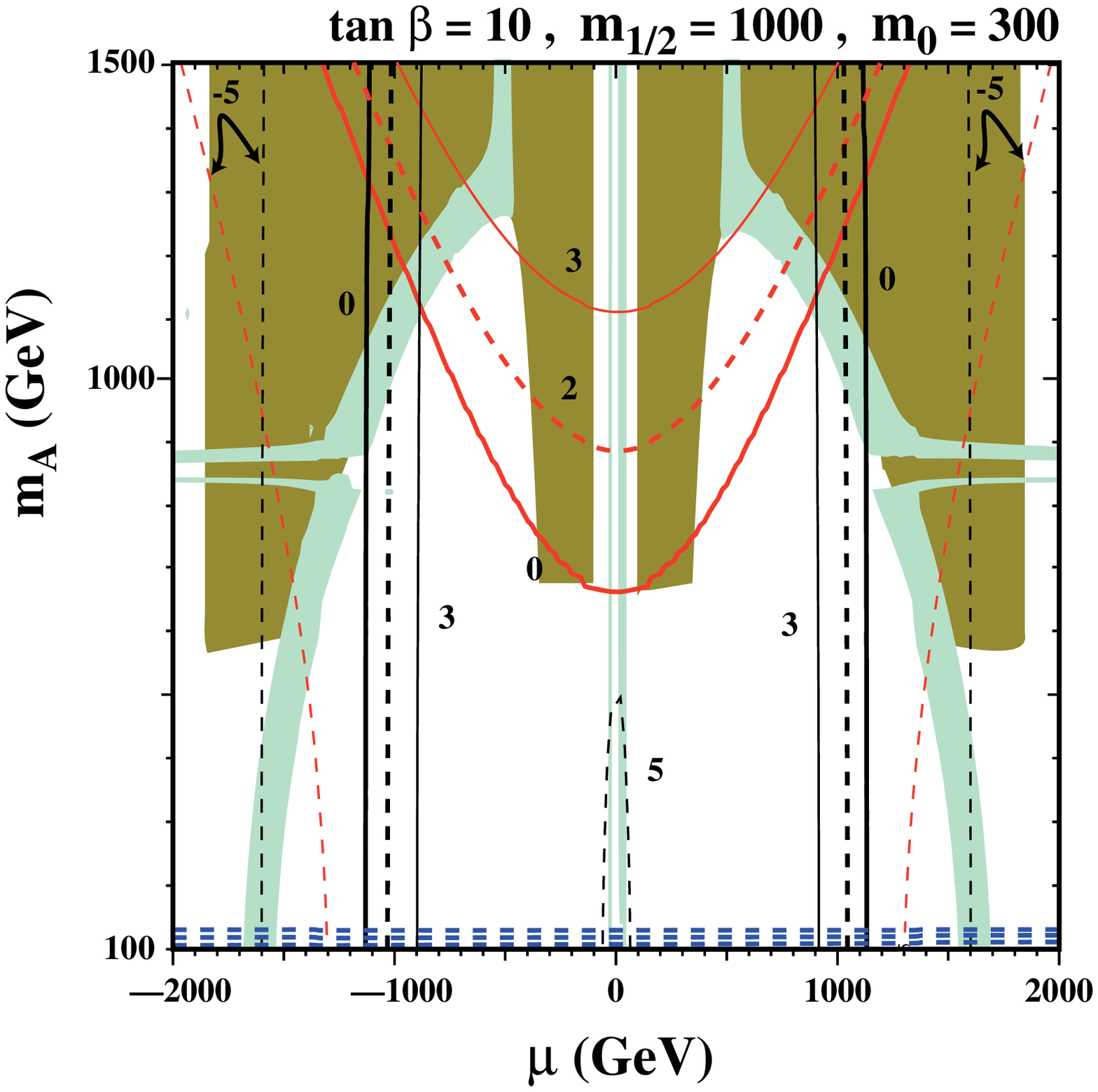,height=3.3in} \hfill
\end{minipage}
\caption{
{\it Contours of the scaled Higgs masses ${\hat m_1}$ and  ${\hat 
m_2}$
in the $(\mu, m_A)$ plane for $\tan \beta = 10$ and (a) $m_0 = 100$~GeV, 
$m_{1/2} = 300$~GeV, (b)  $m_0 = 300$~GeV,
$m_{1/2} = 1000$~GeV, assuming $A_0 = 0$, $m_t = 175$~GeV and 
$m_b(m_b)^{\overline {MS}}_{SM} = 4.25$~GeV.
As in Fig. \protect\ref{fig:mumA10}, the light (turquoise) shading denotes 
the region where
$0.1 <
\Omega_\chi h^2 < 0.3$. The darker shading denotes the region not excluded by
the other constraints.   
The dark (black) lines correspond to contours of  ${\hat m_2}$, and the
lighter (red) lines to contours of  ${\hat m_1}$. In (a) the
thick solid contours correspond to ${\hat m} = 0$, the 
thick dashed contours to ${\hat m} = \pm 5$, and the
thin solid contours to ${\hat m} = -15$.  In (b) the
thick solid contours correspond to ${\hat m} = 0$, the
thick dashed contours to ${\hat m} = 2$, the 
thin solid contours to ${\hat m} = 3$, and the
thin dashed contours to ${\hat m} = \pm 5$.
The dark
(blue) dashed lines at very low values of $m_A$ indicate the contours
$\sin^2 (\beta - \alpha) = 0.7, 0.5$ and $0.3$, which decrease with $m_A$. 
}}
\label{fig:m1m2}
\end{figure}

We next explain the various contours shown in Fig.~\ref{fig:mumA10}.  The
dark (black) dashed line in Fig.~\ref{fig:mumA10}(a) is the contour
$m_{\chi^\pm} = 104$~GeV, representing the LEP kinematic limit on the
chargino mass. The actual LEP lower limit is in fact generally somewhat
smaller than this, depending on the details of the MSSM parameters, but
the differences would be invisible on this plot. We see that the chargino
constraint excludes the $|\mu| \stackrel{<}{\sim} 100$~GeV strip. The
lighter (red) dot-dashed line in Fig.~\ref{fig:mumA10}(a) is the LEP lower
limit of 114~GeV on the mass of the lightest MSSM Higgs boson $h$, as
calculated using the {\tt FeynHiggs} programme \cite{FeynHiggs}. This
limit is relaxed in certain regions of parameter space because of a
suppression of the $ZZh$ coupling but, as discussed below in connection
with Fig.~\ref{fig:m1m2}, this relaxation is irrelevant for the
conclusions presented here. For the choices of the other MSSM parameters
in Fig.~\ref{fig:mumA10}(a), the main effect of the Higgs constraint is
essentially to exclude negative values of $\mu$. The pale (pink) solid
line at $\mu = 0$ marks the $g_{\mu}-2$ constraint. This constraint
excludes the $\mu < 0$ half-plane, while allowing all of the $\mu > 0$
parameter space for this particular choice of $m_{1/2}$ and $m_0$.

The darker (black) dot-dashed line in Fig.~\ref{fig:mumA10}(a) indicates
where one or the other of the Higgs mass-squared becomes negative at the
input GUT scale, specifically when either $(m_1^2 + \mu^2) < 0$ or $(m_2^2
+ \mu^2) < 0$. One could argue that larger values of $|\mu|$ and/or
smaller values of $m_A$ are excluded by requiring the preferred
electroweak vacuum to be energetically favoured and not bypassed early in
the evolution of the Universe. However, for a different point of view,
see~\cite{fors}.

As noted above, the two (black) crosses in Fig.~\ref{fig:mumA10}(a) are
the two pairs
$(\mu, m_A) \simeq (\pm 390, 450)$ GeV obtained in the
CMSSM, assuming that the soft supersymmetry-breaking Higgs scalar masses
are equal to the universal squark and slepton masses-squared at the GUT
input scale (UHM). For this particular choice of the other MSSM parameters
$m_{1/2}, m_0, A_0$ and $\tan \beta$, these two CMSSM points both yield
relic densities
$\Omega_\chi h^2 = 0.21$, within the range preferred by astrophysics and
cosmology. On the other hand, the CMSSM point with $\mu < 0$ has
unacceptable values for $b \to s \gamma$ decay, $a_\mu$,  and $m_h$,
although the latter might conceivably be salvaged if the theoretical
approximations in {\tt FeynHiggs} happen to err in the favourable
direction.

The main conclusions from Fig.~\ref{fig:mumA10}(a) are that moderate
values of $\mu > 0$ are favoured, $m_A$ cannot be small, and there is a
large fraction of the remaining MSSM parameter space where the
cosmological relic density lies within the range favoured by astrophysics
and cosmology. This includes parts of the `bulk' regions identified in
CMSSM studies, with $200~{\rm GeV} \lappeq \mu \lappeq 1000$~GeV and $270~{\rm
GeV} \lappeq m_A \lappeq 650$~GeV, generalizing the CMSSM point for $\mu > 0$.

The notations used for the constraints illustrated in the other panel of
Fig.~\ref{fig:mumA10} are the same, but the constraints interplay in
different ways. In panel (b), we have chosen a larger value of $m_{1/2}$.
In this case, the previous region excluded by the neutral LSP constraint
at large $|\mu|$, migrates to larger $|\mu|$ and is no longer visible in
this panel. However, the `shark's teeth' for moderate $|\mu|$ grow,
reaching up to $m_A \sim 1250$~GeV. These arise when one combines a large
value of $m_{1/2}$ with a relatively small value of $m_0$, and one may
find a $\stau$ or even a $\tilde e$ LSP.  The large value of $m_{1/2}$
also keeps the rate of $b \to s \gamma$ under control unless $m_A$ is very
small.  The chargino constraint is similar to that in panel (a), whereas
the $m_h$ constraint is irrelevant due to the large value of $m_{1/2}$.
The $g_\mu - 2$ constraint does not provide any exclusion here.  Finally,
we observe that the GUT mass-squared positivity constraint now allows
larger values of $|\mu| \lappeq 1800$~GeV.  In this example, the two CMSSM
points are at $(\mu, m_A) \simeq (\pm 1100, 1330)$ GeV and both have relic
densities that are too large: $\Omega_\chi h^2 \simeq 1.15$.

The main conclusions from Fig.~\ref{fig:mumA10}(b) are that moderate 
values of
$\mu$ are favoured, which may be of either sign, but still $m_A$ cannot be
small. As in panel (a), there is a large fraction of the remaining MSSM
parameter space where the cosmological relic density lies within the range
favoured by astrophysics and cosmology, for both signs of $\mu$. 

The reader may be wondering by now: how non-universal are the Higgs masses
in the previous plots? Do they differ only slightly from universality - in
which case the usual CMSSM results would be very {\it unstable}
numerically, or do they involve violations of universality by orders of
magnitude - in which case the NUHM discussion would be unimportant and the
usual CMSSM results would be very {\it stable} numerically? Some answers
are provided in Fig.~\ref{fig:m1m2}, where we plot contours of ${\hat m_1}
\equiv {\rm sign}(m_1^2) \times |m_1 / m_0|$ as paler (red) curves and of
${\hat  m_2} \equiv
{\rm sign}(m_2^2) \times |m_2 / m_0|$ as darker (black) curves. We see in
panel  (a), for
$\tan \beta = 10$, $m_{1/2} = 300$, $m_0 = 100$~GeV and $A_0 = 0$,
corresponding to panel (a) of Fig.~\ref{fig:mumA10}, that relatively large
values are attained for both ${\hat m_1}$ and ${\hat m_2}$, ranging up to
15 or more in modulus.  However, comparing Fig.~\ref{fig:mumA10}(a) and
Fig.~\ref{fig:m1m2}(a), we see that such large values are not attained in
the restricted region of the $(\mu, m_A)$ plane that are allowed by the
various phenomenological constraints discussed earlier. We have 
indicated by darker shading the
regions of Fig.~\ref{fig:m1m2}(a) which are allowed by the
non-cosmological constraints, and see that $|{\hat m_{1,2}}| \lappeq 5$ in
the regions allowed.

Turning now to panel (b) of Fig.~\ref{fig:m1m2}, for $\tan \beta = 10$,
$m_{1/2} = 1000$, $m_0 = 300$~GeV and $A_0 = 0$, we see values of $|{\hat
m_{1,2}}| \lappeq 5$ throughout the portion of the plane displayed.
Comparing again with the corresponding panel (b) of Fig.~\ref{fig:mumA10},
we see that most of the displayed range of $\mu$ is allowed by the
experimental constraints, but only for $m_A \gappeq 600$~GeV.  In this region,
we find $|{\hat m_{1,2}}| \lappeq 5$.

We conclude that the answer to the questions posed earlier lie in between
the extremes proposed. The variations in $|{\hat m_{1,2}}|$ are by no
means negligible, but neither are they excessive within the regions of
parameter space allowed by the experimental constraints. The CMSSM results
are not excessively unstable, but significant variations are possible for 
plausible ranges of ${\hat m_{1,2}}$.

We comment finally on the magnitude of the $ZZh$ coupling $\sin^2 (\beta -
\alpha)$, which controls the observability of the lightest MSSM Higgs
boson at LEP. The contours $\sin^2 (\beta - \alpha) = 0.7, 0.5$ and $0.3$
are shown as closely-spaced dashed lines at the bottom of each panel in 
Fig.~\ref{fig:m1m2}. In the bulk
of the $(\mu, m_A)$ planes shown, the $ZZh$ coupling is close to its
Standard Model value, and the LEP lower limit on $m_h$ is
indistinguishable from the Standard Model value limit of 114~GeV.

\section{Analysis of the $(\mu, M_2)$ Plane}

We now turn to projections of the MSSM parameter space on the $(\mu, M_2)$
plane, which is often used in the analysis of the chargino and neutralino
sectors of the MSSM. In the past, it has been demonstrated how the
cosmological constraint on the relic neutralino density may be satisfied
in a large part of the $(\mu, M_2)$ plane, for certain values of the other
MSSM parameters~\cite{nonu,higgsino,EFGOS,EFGO}. Here we limit ourselves
to a couple of examples  that indicate how the cosmological region may
vary.

\begin{figure}
\vskip 0.5in
\vspace*{-0.75in}
\begin{minipage}{8in}
\epsfig{file=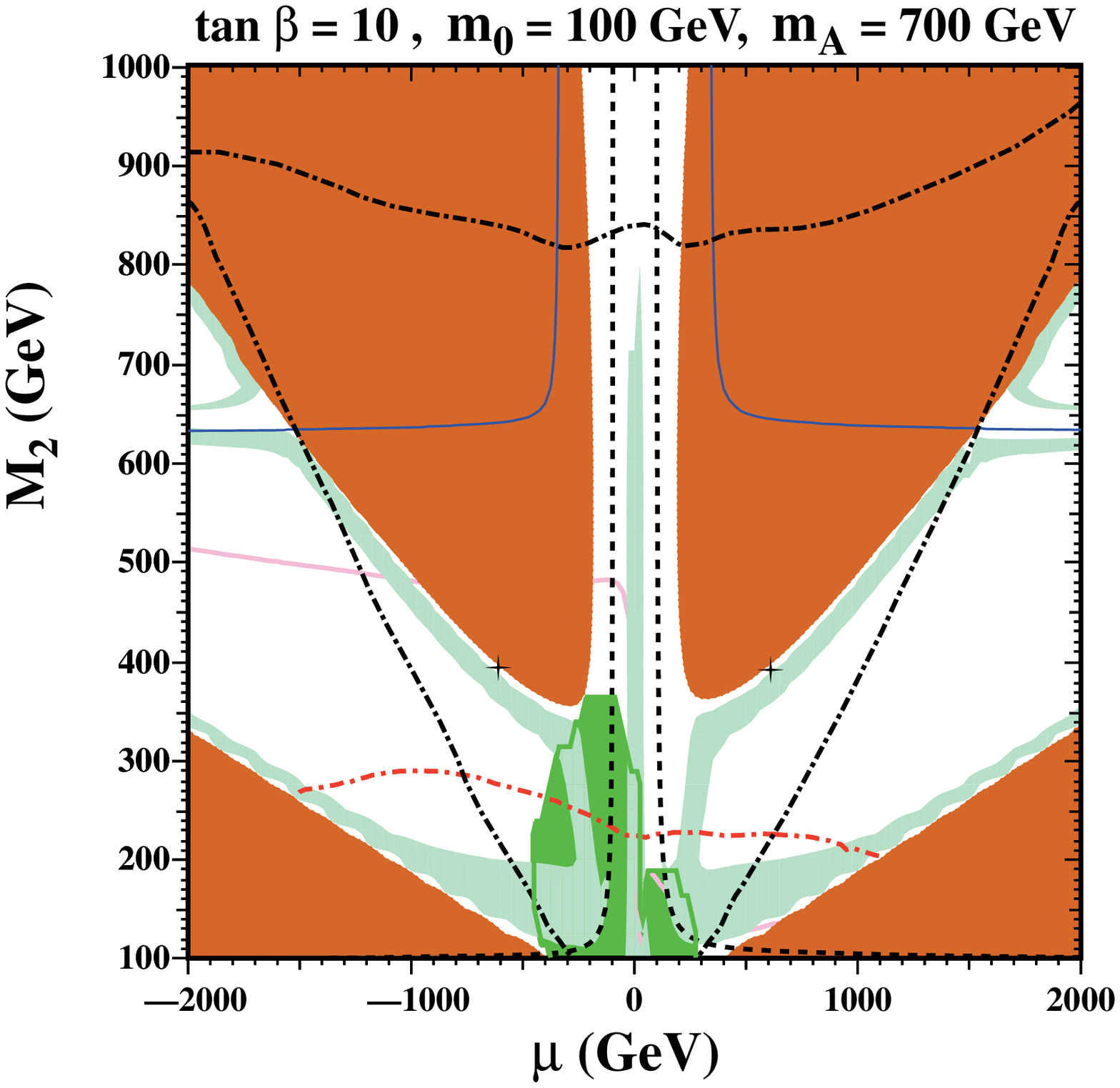,height=3.3in}
\hspace*{-0.17in}
\epsfig{file=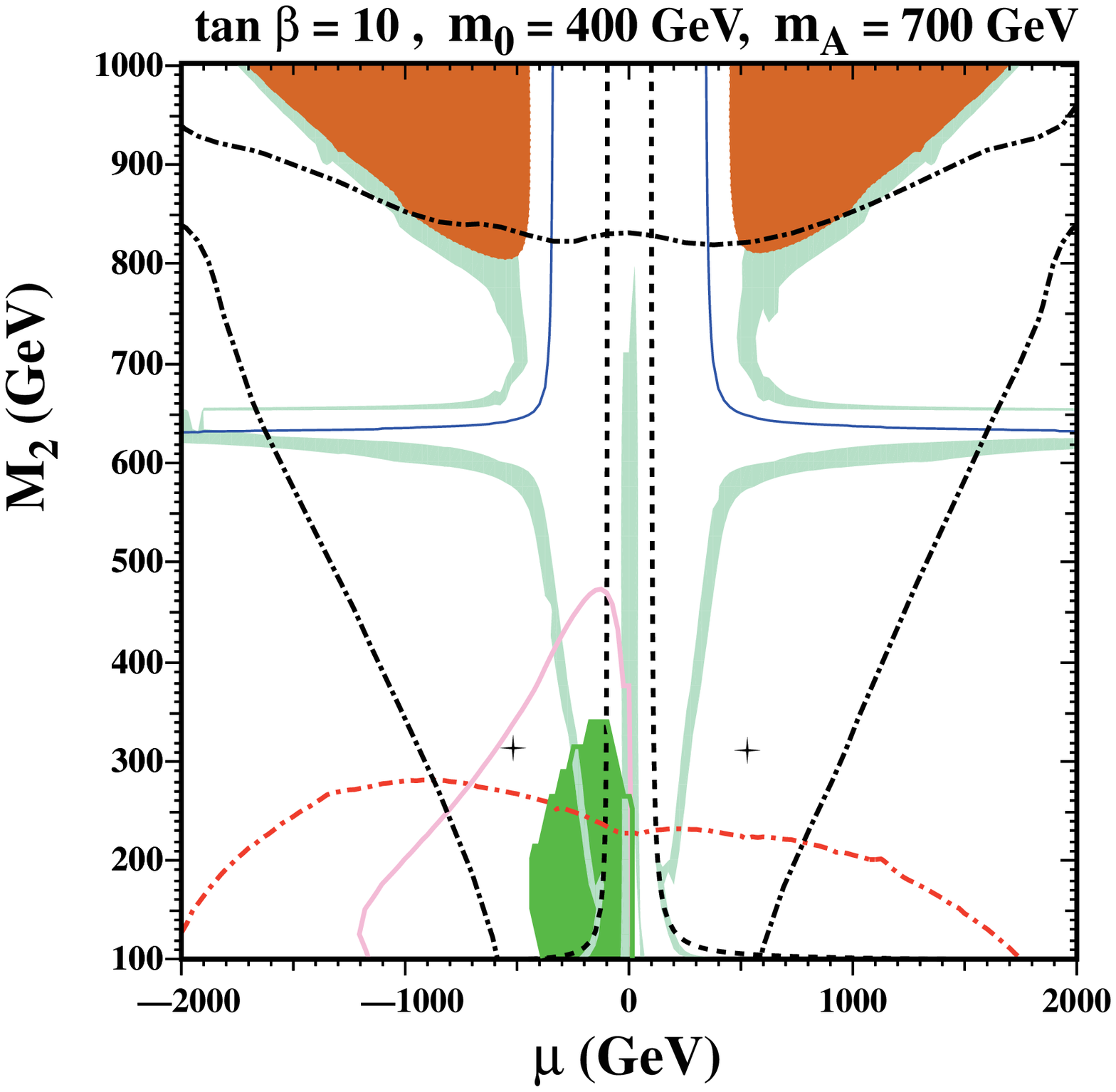,height=3.3in} \hfill
\end{minipage}
\caption{
{\it 
Compilations of phenomenological constraints on the MSSM with NUHM
in the $(\mu, M_2)$ plane for $\tan \beta = 10$ and (a) $m_0 = 100$~GeV,
$m_A = 700$~GeV, (b)  $m_0 = 400$~GeV,
$m_A = 700$~GeV, again assuming $A_0 = 0$, $m_t = 175$~GeV and
$m_b(m_b)^{\overline {MS}}_{SM} = 4.25$~GeV.
The pale (turquoise) shading denotes the region where $0.1 < \Omega_\chi 
h^2 < 0.3$, and  
the (blue) solid line is the contour $m_\chi = m_A/2$, near which
rapid direct-channel annihilation suppresses the relic density.
The darker (green) shading shows the impact of the $b \to s \gamma$
constraint, and the darkest (red) shading shows where the LSP is charged.
The dark (black) dashed line is the chargino constraint $m_{\chi^\pm} > 
104$~GeV: lower values of $|\mu|$ and/or $M_2$ are not allowed.
The lighter (red) dot-dashed line is the contour $m_h = 114$~GeV
calculated using {\tt FeynHiggs}~\cite{FeynHiggs}: lower $m_A$ are not 
allowed. The pale (pink) solid line shows the region excluded by $g_\mu - 
2$.
The dark (black) dot-dashed triangular line indicates when one or another 
Higgs
mass-squared becomes negative at the GUT scale: only lower $|\mu|$ and
intermediate $m_{1/2}$ are allowed. The (black) crosses denote the CMSSM points.
}}
\label{fig:muM2} 
\end{figure}

Panel (a) of Fig.~\ref{fig:muM2} displays the $(\mu, M_2)$ plane for the
choices $\tan \beta = 10$, $m_0 = 100~{\rm GeV}$, $m_A = 700~{\rm GeV}$
and $A_0 = 0$. In this case, the region favoured by cosmology, shown by
the light (turquoise) shading, is in the part of the plane where the LSP
$\chi$ is mainly a Bino, as preferred in the CMSSM.  The LEP chargino
constraint, shown as a dark (black) dashed line, excludes small values of
$\mu$ and/or $M_2$, where a Higgsino LSP might have constituted the dark
matter.  The $m_h$ constraint, shown as a paler (red) dot-dashed line,
excludes low values of $M_2$, particularly for $\mu < 0$, but allows
substantial fractions of the cosmological regions. The $b \to s \gamma$
constraint, shown in darker shading, excludes another part of the
remaining allowed region for $\mu < 0$, but leaves almost untouched the
allowed region for $\mu > 0$. The $g_\mu - 2$ constraint, shown as a pale
(pink) solid line, excludes a larger region of $\mu < 0$, but leaves an
allowed region at higher $M_2$. The requirement that the effective Higgs
masses-squared be positive at the GUT scale allows a triangular region
centred around $\mu = 0$ and extending up to $M_2 \sim 900$~GeV, bounded
by the dark (black) dot-dashed lines, and is compatible with all the other
constraints in regions with both signs of $\mu$. The cosmological region
is bounded above by the dark (red) shaded region where the LSP is charged,
close to which coannihilation is important in suppressing the relic
density to an acceptable level. Also shown at large $M_2$ is the solid
(blue) line where $m_\chi = m_A/2$, but this has little effect on the
relic density in the region allowed by the other constraints.

Panel (b) of Fig.~\ref{fig:muM2} displays the $(\mu, M_2)$ plane for the
choices $\tan \beta = 10, m_0 = 400~{\rm GeV}$, $m_A = 700~{\rm GeV}$ and
$A_0 = 0$. In this case, the cosmological region is largely complementary
to panel (a), since both the Bino and Higgsino regions are excluded. Only
narrow strips in the regions where the LSP is a strong mixture of gaugino
and Higgsino have an acceptable relic density, with the exception of
indentations where $m_\chi \sim m_A / 2$ and broader regions at large
$M_2$, where contributions from $s$-channel annihilation and $\tilde \tau$
coannihilation are both important. The chargino, Higgs, $b \to s \gamma$,
$g_\mu - 2$ and GUT positivity constraints interplay much as in panel (a).

We have also examined the $(\mu, M_2)$ planes for other choices of $m_0$
and $m_A$. With a judicious choice of these parameters, a large fraction
of the domain where $\mu \gappeq M_2$ and the LSP is gaugino-like may
happen to have an acceptable relic density, though it may then be excluded
by other constraints. However, we have not found any instance where a
mainly Higgsino-like LSP is permitted with a sufficiently large relic
density:  see also~\cite{EFGOS}. This is largely due to the LEP lower
limit on $m_{\chi^\pm}$ and the fact that large $M_2$ is excluded at small
$|\mu|$ by the neutral LSP requirement.

\section{Analysis of the $(m_{1/2}, m_0)$ Plane}

This projection of the MSSM parameter space has often been used in studies
of supersymmetric dark matter, in particular in the context of the CMSSM, 
where it was instrumental in the specification~\cite{benchmark} of 
benchmark scenarios 
compatible with all the experimental and cosmological constraints 
discussed earlier. Here we discuss examples in the more general MSSM 
context, which indicate some of the range of new possibilities that it 
opens up.

\begin{figure}
\vskip 0.5in
\vspace*{-0.75in}
\begin{minipage}{8in}
\epsfig{file=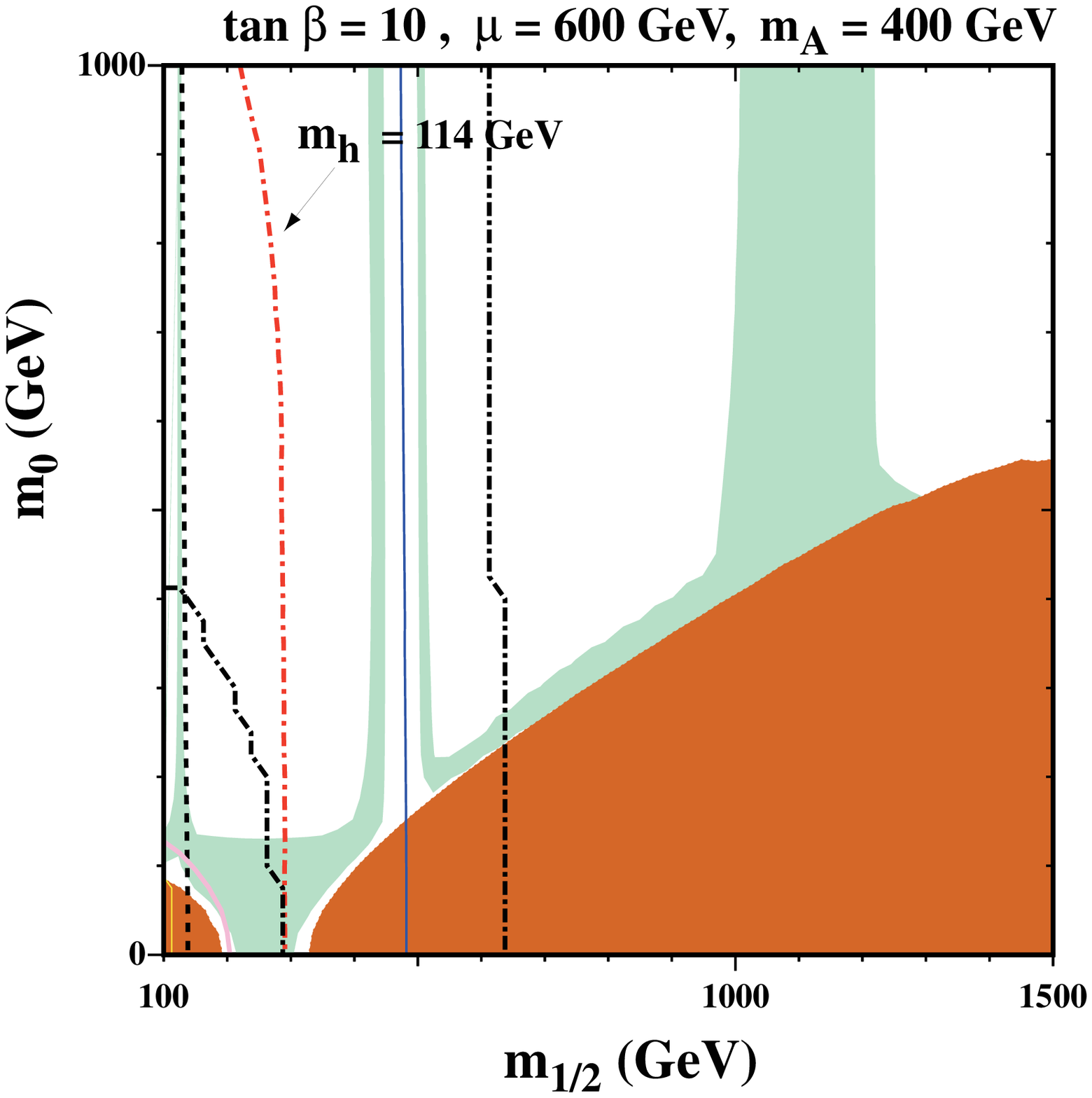,height=3.3in}
\hspace*{-0.17in}
\epsfig{file=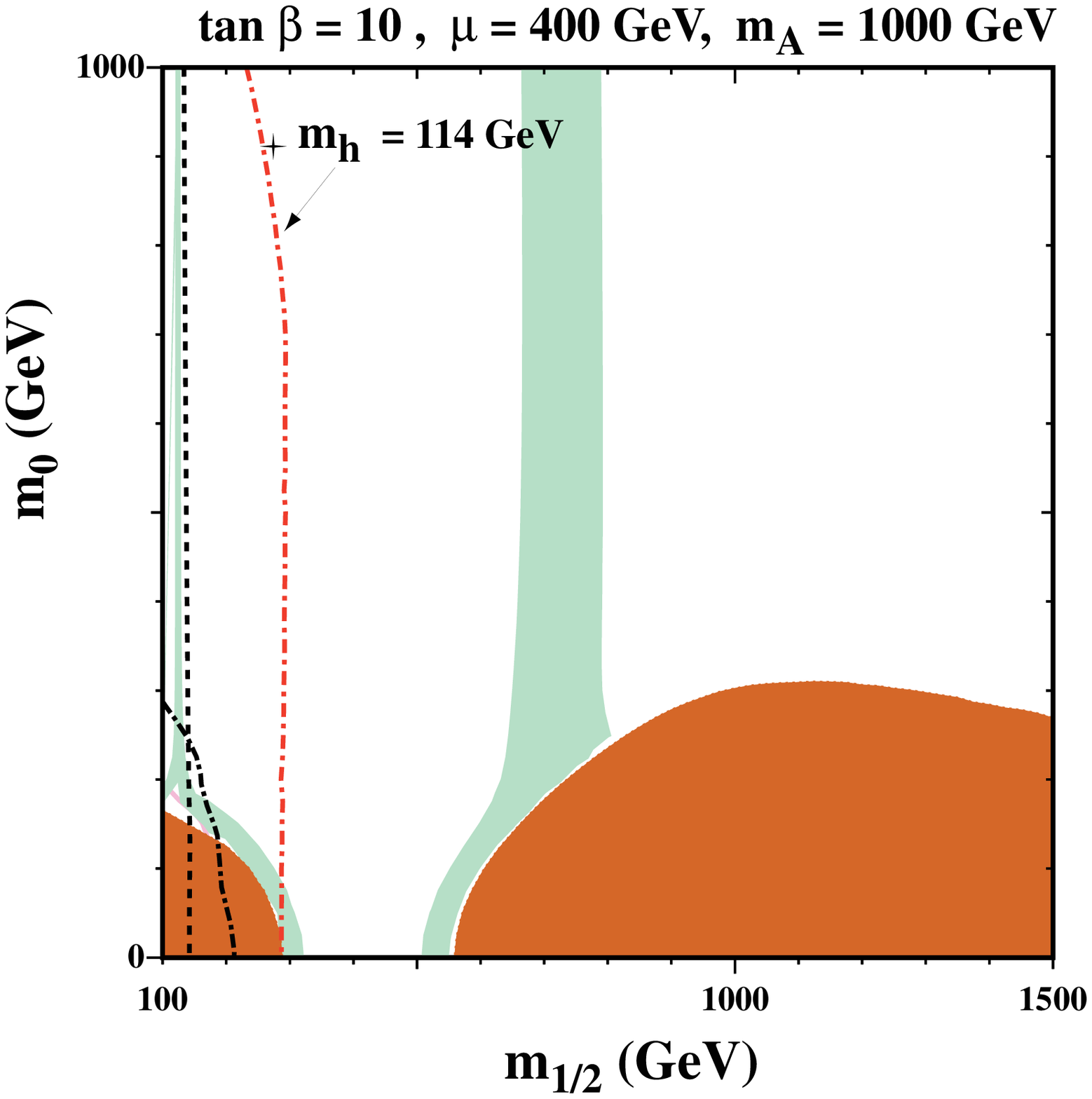,height=3.3in} \hfill
\end{minipage}
\caption{
{\it
Compilations of phenomenological constraints on the MSSM with NUHM
in the $(m_{1/2}, m_0)$ plane for $\tan \beta = 10$ and (a) $\mu = 
600$~GeV, $m_A = 400$~GeV, (b) $\mu = 400$~GeV,
$m_A = 1000$~GeV, again assuming $A_0 = 0$, $m_t = 175$~GeV and
$m_b(m_b)^{\overline {MS}}_{SM} = 4.25$~GeV.
The pale (turquoise) shading denotes the region where $0.1 < \Omega_\chi 
h^2 < 0.3$, and  
the (blue) solid line is the contour $m_\chi = m_A/2$, near which
rapid direct-channel annihilation suppresses the relic density.
The darkest (red) shading shows where the LSP is charged.
The dark (black) dashed line is the chargino constraint $m_{\chi^\pm} > 
104$~GeV: lower $m_{1/2}$ are not allowed.
The lighter (red) dot-dashed line is the contour $m_h = 114$~GeV
calculated using {\tt FeynHiggs}~\cite{FeynHiggs}: lower $m_{1/2}$ are not 
allowed.
The light (pink) solid line is where the supersymmetric contribution to the
muon anomalous moment is $a_\mu = 58 \times 10^{-10}$: lower $m_0$ and $m_{1/2}$
are excluded at the $2-\sigma$ level. 
The dark (black) dot-dashed lines indicates when one or another Higgs
mass-squared becomes negative at the GUT scale: only intermediate values 
of $m_{1/2}$ are allowed in panel (a), and larger values in (b). The 
(black) cross in panel (b) denotes the CMSSM
point.
}}
\label{fig:m12m0}
\end{figure}

Panel (a) of Fig.~\ref{fig:m12m0} shows the $(m_{1/2}, m_0)$ plane for
$\tan \beta = 10$ and the particular choices $\mu = 600$~GeV and $m_A =
400$~GeV. The dark (red)  shaded regions are excluded because the LSP is
charged: the larger part resembles the similar excluded regions in the
CMSSM. As in the CMSSM studies, there are light (turquoise) shaded strips
close to these forbidden regions where coannihilation suppresses the relic
density sufficiently to be cosmologically acceptable. Further away from
these regions, the relic density is generally too high. The near-vertical
dark (black) dashed and light (red) dot-dashed lines are the LEP exclusion
contours
$m_{\chi^\pm} > 104$~GeV and $m_h > 114$~GeV. As in the CMSSM case, they
exclude low values of $m_{1/2}$, and hence rule out rapid relic
annihilation via direct-channel $h$ and $Z^0$ poles.

A striking feature when $m_{1/2} \sim 400$ to $500$~GeV is a `funnel' with
a double strip of acceptable relic density. This is due to rapid
annihilation via the direct-channel $A, H$ poles which occur when $m_\chi
= m_A / 2 = 200$~GeV, indicated by the solid (blue) line. Inside the
shaded double strip, the funnel contains an unshaded strip where the relic
density falls below the range preferred by cosmology in the absence of
other types of cold dark matter. The existence of analogous
rapid-annihilation funnels  has been noticed previously in the CMSSM
context: there they were diagonal  in the $(m_{1/2}, m_0)$ plane, because
the CMSSM imposes a link between 
$m_0$ and $m_A$ that is absent in the more general MSSM discussed here.

There is also another strip in Fig.~\ref{fig:m12m0}(a)  around $m_{1/2}
\sim 1100$~GeV where the relic density falls again into the allowed range.
In this region, the neutralino acquires enough Higgsino content for the
relic density be in the range preferred by cosmology. For larger
$m_{1/2}$, the relic density is suppressed even more by $\chi -
\chi^\prime - \chi^\pm$ coannihilation, and for $m_{1/2}
\stackrel{>}{\sim} 1200$~GeV the relic density falls down below 0.1.
This strip along with most of the $(m_{1/2}, m_0)$ plane shown in
Fig.~\ref{fig:m12m0}(a) is actually excluded by the requirement that the
Higgs scalar masses be positive at the input GUT scale, as indicated by
the dark (black) dot-dashed lines. However, the rapid-annihilation funnel
is still allowed by this constraint, as is a part of the $\chi - {\tilde
\ell}$ coannihilation region. The CMSSM point is located beyond the region
shown here.

Panel (b) of Fig.~\ref{fig:m12m0} shows the $(m_{1/2}, m_0)$ plane for
$\tan \beta = 10$ and the different choices $\mu = 400$~GeV and $m_A =
1000$~GeV. In this case, the dark (red) shaded charged-LSP region has
rather different shape, but still excludes a substantial region with large
$m_{1/2}$ and small $m_0$.  
The striking feature of Fig.~\ref{fig:m12m0}(b) is the broad band of
allowed relic density around $m_{1/2} \sim 700$~GeV, where the relic
density is suppressed into the preferred range by $\chi - \chi^\prime -
\chi^\pm$ coannihilation, the relic density falling below 0.1 for $m_{1/2}
\stackrel{>}{\sim} 800$~GeV. In this case, the region allowed by the GUT
stability requirement extends up to $m_{1/2} \sim 1600$~GeV, and is
therefore satisfied throughout most of the displayed part of the
$(m_{1/2}, m_0)$ plane.

In the small $m_0 \stackrel{<}{\sim} 200$~GeV and $m_{1/2}
\stackrel{<}{\sim} 300$~GeV corner, the tau sneutrino can become lighter
than the stau, due to the negative value of $m_2^2 - m_1^2$. Including
$\chi - \tilde{\nu}_{\tau}$ coannihilation would shift the region 
preferred by cosmology for the $\chi$ LSP to somewhat higher 
values of $m_0$ and $m_{1/2}$. Note that, in 
a very narrow strip between the $\chi$- and ${\tilde \tau}$-LSP regions,
the LSP is in fact ${\tilde \nu_\tau}$.  The direct-channel $H, A$
rapid-annihilation region is not seen in panel (b), because the
neutralino has already become Higgsino-like with a mass
$\sim 400$~GeV when $m_{1/2}
\sim 1000$~GeV, and therefore $m_{\chi}$ is always below $m_A/2$. For
other MSSM choices, the rapid-annihilation region may overlap with the
broad band where $\chi - \chi^\prime - \chi^\pm$ coannihilation begins to
become important.
 
These two examples serve to demonstrate that the $(m_{1/2}, m_0)$ plane
may look rather different in the CMSSM from its appearance in the CMSSM
for the same value of $\tan \beta$. In particular, the locations of
rapid-annihilation funnels and $\chi - \chi^\prime - \chi^\pm$
coannihilation regions are quite model-dependent, and the GUT stability
requirement may exclude large parts of the $(m_{1/2}, m_0)$ plane.

\section{Analysis of the $(m_A, \tan \beta)$ Plane}

\begin{figure}
\vskip 0.5in
\vspace*{-0.75in}
\begin{minipage}{8in}
\epsfig{file=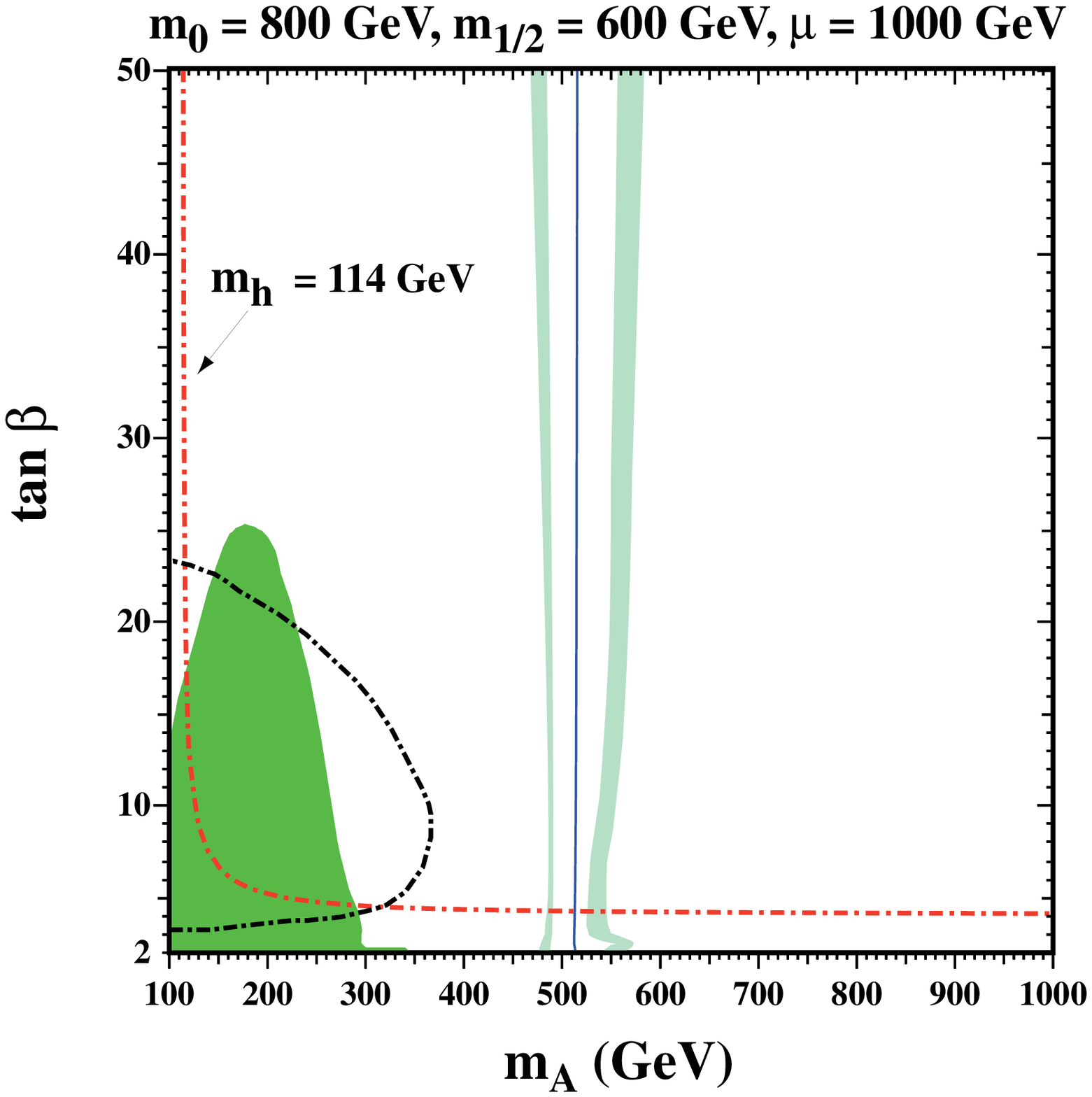,height=3.3in}
\hspace*{-0.17in}
\epsfig{file=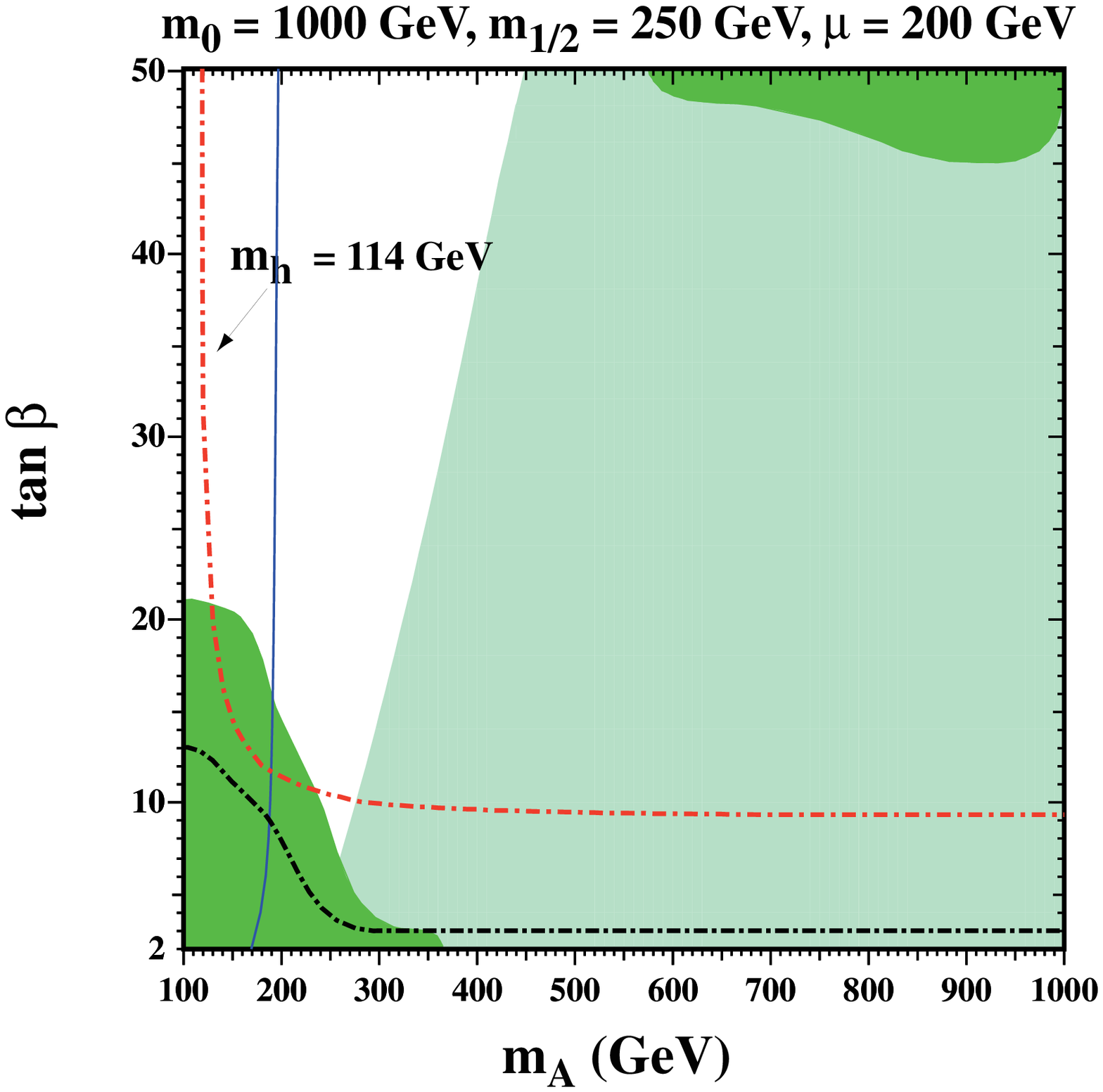,height=3.3in} \hfill
\end{minipage}
\caption{
{\it Compilations of phenomenological constraints on the MSSM
with NUHM    in the $(m_A, \tan \beta)$ plane for $(m_{1/2}, m_0, \mu) = 
{\rm (a)} (600, 800, 1000), {\rm (b)} (250, 1000, 200)$~GeV.
The lighter (red) dot-dashed lines are the 
contours $m_h = 114$~GeV,
the solid (blue) lines show where $m_\chi \sim m_A / 2$
and the dark (black) dot-dashed line indicates when one or another Higgs
mass-squared becomes negative at the GUT scale. The light 
(turquoise) shading indicates where $0.1 < \Omega_\chi h^2 < 0.3$, the 
darker (green) shaded regions are excluded by the $b \to s \gamma$ 
constraint.
}}
\label{fig:mAtb}
\end{figure}

This projection of the MSSM parameter space is often used in discussions
of Higgs phenomenology, and a number of benchmark Higgs scenarios have
been proposed~\cite{CHWW}. As we now see in more detail, these do not
always take fully into account other phenomenological constraints on the
MSSM. We give examples of $(m_A, \tan \beta)$ planes for selected values
of the other MSSM parameters in Fig.~\ref{fig:mAtb}. We first note the
following general features. The LEP constraint on $m_h$ excludes a region
at low $m_A$ and/or $\tan \beta$, and the $b \to s \gamma$ constraint also
removes large domains of the $(m_A, \tan \beta)$ planes, whose locations
depend on the other MSSM parameters. The requirement that the Higgs
masses-squared be positive at the GUT scale also excludes large domains of
parameter space. However, the most striking constraint is that imposed by
the cosmological relic density.

We discuss first panel (a) of Fig.~\ref{fig:mAtb}, for $(m_{1/2}, m_0,
\mu) = (600, 800, 1000)$~GeV. This is similar to one of the Higgs
benchmark scenarios proposed in~\cite{CHWW}, which has $\mu = 2000$~GeV.
However, we find that, for such a large value of $\mu$, the GUT positivity
constraints would be disobeyed all over the $(m_A, \tan \beta)$ plane.
This reflects the fact the good renormalization-group running up to the 
GUT scale was not
considered as a criterion in selecting the Higgs benchmark points. 
Reducing $\mu$ to the value $1000$~GeV shown in panel (a) of 
Fig.~\ref{fig:mAtb}, the GUT positivity
constraints exclude a region at small $m_A$ and $\tan \beta$, which is
largely excluded also by the $b \to s \gamma$ constraint. In this panel,
there is only a very narrow range of values of $m_A$ where the relic
density falls within the range favoured by cosmology. It corresponds to
one of the `funnels' noted previously in an analysis of the CMSSM at large
$\tan \beta$~\cite{EFGOSi}, and is actually divided into two strips. The
broader strip of $m_A$ values is just above $m_A \sim 2 m_\chi$, and a
narrower strip appears just below $m_A \sim 2 m_\chi$. In between, there
is a narrow strip of $m_A$ values where the relic density lies below the
preferred cosmological range: this strip would be allowed if there is
another important source of cold dark matter. Outside the double strip
around $m_A \sim 2 m_\chi$, the relic density is higher than allowed by
cosmology. This exclusion could be evaded only by postulating that the
lightest neutralino is unstable, either in an $R$-conserving model if
there is a lighter sparticle such as the gravitino, or in an $R$-violating
model. Thus most of the plane in panel (a) of Fig.~\ref{fig:mAtb} is
excluded, even after reducing $\mu$.

The picture changes strikingly in panel (b) of Fig.~\ref{fig:mAtb}, for
the choices $(m_{1/2}, m_0, \mu) = (250, 1000, 200)$~GeV. The contour $m_h
= 114$~GeV is shown as a lighter (red) dot-dashed line descending steeply
when $m_A \sim 130$~GeV, and then flattening out at $\tan \beta \sim 10$.  
The vertical solid (red) line shows where $m_\chi \sim m_A / 2$. The dark
(green) shaded regions at $(m_A, \tan \beta) \lappeq (300~{\rm GeV}, 20)$
and $ \gappeq (600~{\rm GeV}, 45)$ are excluded by the $b \to s \gamma$
constraint. We see in this case that the relic density constraint,
indicated by the light (turquoise) shading, is satisfied throughout a
broad swathe of $m_A \gappeq 250$~GeV for $\tan \beta = 5$ to $m_A \gappeq
450$~GeV for $\tan \beta = 50$. The requirement that the Higgs
masses-squared be positive at the GUT scale excludes a domain of
parameter space at low $\tan \beta$, as indicated by the dark dot-dashed 
line.

The Higgs benchmark scenarios were not chosen with the cosmological relic
density in mind, whereas this was taken explicitly into account in
formulating the sparticle benchmark scenarios proposed
in~\cite{benchmark}.  It would be interesting to study NUHM Higgs
benchmark scenarios that respect the cosmological relic density constraint
more systematically. This could perhaps be done by postulating a value of
$m_{1/2}$ that varies with $\tan \beta$.  However, a detailed study of
this point goes beyond the scope of this paper.

\section{Conclusions}

We have shown in this paper that relaxing the scalar-mass universality
assumption for the MSSM Higgs multiplets opens up many phenomenological
possibilities that were not evident in CMSSM studies with universal
masses. We have emphasized the importance of requiring the LSP to be
neutral and imposing the positivity of scalar masses-squared at the GUT
scale. We find that a mainly Bino neutralino LSP is still preferred,
though it may also be mixed with a large Higgsino component. However, we
do not find MSSM parameter regions where the LSP is mainly a Higgsino. In
addition to direct-channel $A, H$ poles and $\chi - {\tilde \ell}$
coannihilation, we have identified generic instances where $\chi -
\chi^\prime - \chi^\pm$ coannihilation is important. We have also shown 
that Higgs benchmark scenarios do not respect in general the full range of 
phenomenological requirements. It is desirable to look for an agreed set 
of MSSM benchmarks that incorporate these in the studies of Higgs 
phenomenology, which we do not attempt here.

The higher-dimensional parameter space of the MSSM with non-universal 
Higgs masses is clearly much richer than the simplified CMSSM, and we have 
only been able to touch on a few of the more striking aspects in this 
paper. More work is needed to digest more fully the impacts of the 
different experimental, phenomenological, theoretical and cosmological 
constraints. There are surely many more interesting features beyond those 
mentioned here.

\section*{Acknowledgments}

We thank Toby Falk for many valuable discussions.
The work of  K.A.O. and Y.S. was supported partly by DOE grant
DE--FG02--94ER--40823.

\end{document}